\documentclass[aps,pra,twocolumn,draft,groupedaddress,amsmath,eqsecnum,amssymb]{revtex4}

\begin{document}


\title{Open environments for quantum open systems}


\author{Michael R. Gallis}
\email[]{mrg3@psu.edu}
\homepage[]{http://phys23p.sl.psu.edu/~mrg3}
\affiliation{Capital College, Penn State University, Schuylkill
Campus, 200 University Drive, Schuylkill Haven, Pennsylvania
17972}

\date{\today}

\begin{abstract}
The majority of quantum open system models in the literature are
simplistic in the sense that they only explicitly account for that
part of the environment that directly interacts with the system of
interest. A quantum open system with an open environment is
examined using the projection operator method in the weak coupling
limit. The openness of environment is modelled by nonunitary
evolution of the Lindblad form. Under certain conditions, the
resulting master equation for the system is insensitive to the
initial state of the environment and to initial entanglements
between the system and environment for time scales greater than
the environment relaxation timescales. For the particular case of
an environment consisting of a harmonic oscillator bath, the
resulting master equations are demonstrated to have the algebraic
form for completely positive evolution.  The open environment
model is illustrated for the particular case of a system linearly
coupled to an oscillator bath.
\end{abstract}

\maketitle

\section{Introduction}

There has been much interest in quantum open systems (particularly
with applications to quantum decoherence
\cite{Zurek1,Zurek2,Giulini+} over the past several
decades.\cite{Alicki3}

As discussed by Alicki,\cite{Alicki3} models that have been
extensively studied in the literature have arisen from one of two
approaches. The first starts with an \textit{a priori} proposal of
a mathematical form for the quantum dynamical semigroup (Alicki's
\textit{axiomatic approach}), and proceeds with a study of the
resulting properties of the dynamics.  The second is to study the
reduced dynamics of the system of interest which is part of a
composite system which also includes environmental degrees of
freedom (Alicki's \textit{constructive approach}).  The models
primarily explored for the constructive approach have a set of
common features. The system of interest evolves unitarily in
isolation. The environment (often an oscillator bath) evolves
unitarily in isolation. The interaction between system and
environment consists of a term in the composite hamiltonian, so
that the composite system evolves unitarily. The initial state of
the composite system is an important feature of the particular
model being studied and is often but not exclusively taken to be a
factor state between an initial state of the system of interest
and a thermal equilibrium state for the environment.

The study of open systems is an attempt to describe how ``the rest
of the universe" influences the system of interest. However, most
constructive models only include that part of the rest of the
universe with which the rest of the universe directly interacts
(through an interaction term in the composite hamiltonian). In
order to account for that part of the universe with which the
system of interest does not directly interact, we explore
properties of models for which the environment degrees of freedom
are themselves open systems.  In the Section~\ref{class}, we
motivate the quantum models by examining a classical system and
environment where the environment degrees of freedom are damped
oscillators relaxing into thermal equilibrium. In
Section~\ref{env_prop} we review some important properties of the
model we will use to simulate the open nature of the environment.
In Section~\ref{proj} we proceed to use the projection operator
method in the weak coupling regime on a composite system where the
environment relaxes to thermal equilibrium via a nonunitary
evolution of the Lindblad form.\cite{Lindblad}  We illustrate the
results for the particular case of linear coupling to an
oscillator bath in Section~\ref{ex}. We finish in
Section~\ref{done} with a discussion of our results.

\section{\label{class}Classical Open System}

In this section we examine a classical open system with a damped
environment to gain insight and anticipate properties of analogous
quantum models.  We are essentially following
Zwanzig\cite{zwanzig} with the addition of dissipation and
thermalization to the dynamics of the environment.

The system is described in terms of its coordinate $x$.  The
environment consists of a bath of independent oscillators with
coordinates $q_{\mu}$.  The composite system equations of motion
are given by
\begin{equation}
m \ddot{x} = -\frac{\partial U(x)}{\partial x} +\sum_{\mu}
m_{\mu}\omega_{\mu}^{2} [ q_{\mu}-a_{\mu}(x) ] \frac{\partial
a_{_\mu}(x)}{\partial x}
 \label{cde:1}
\end{equation}
for the system coordinate, and
\begin{eqnarray}
m_{\mu} \ddot{q_{\mu}} &=& -m_{\mu}\omega_{\mu}^{2}q_{\mu} - 2
m_{\mu}\gamma_{\mu}\dot{q}_{\mu} \nonumber\\
&&+m_{\mu}\omega_{\mu}^{2} a_{\mu}(x) +F_{\mu}(t) \label{cde:2}
\end{eqnarray}
for each environment degree of freedom.  These equations of motion
correspond to the addition of dampening (with friction
coefficients $ \eta_{\mu} = 2 m_{\mu} \gamma_{\mu} $ ) and thermal
noise terms $ F_{\mu} $ to the environment portion of the closed
system dynamics obtained from the composite system Lagrangian:
\begin{equation}
L=\frac{1}{2} m \dot{x}^{2} - U(x)+ \sum_{\mu} \frac{m_{\mu}}{2}
\{\dot{q}_{\mu}^{2}-\omega_{\mu}^{2} [ q_{\mu}-a_{\mu}(x) ]^{2}\}
 \label{cde:3}
\end{equation}
Each oscillator is taken to be independent but driven to the same
temperature by dissipation and white noise so that the following
fluctuation-dissipation relation holds:
\begin{equation}
\langle F_{\mu}(t)F_{\nu}(s) \rangle =
4\gamma_{\mu}m_{\mu}k_{B}T\delta_{\mu\nu}\delta(t-s).
 \label{cde:efd}
\end{equation}
The Kronecker delta indicates the oscillators in the bath are
driven by independent noise terms.  The Dirac delta function
indicates that the driving forces are white noise.

With $\Omega_{\mu}\equiv \sqrt{\omega_{\mu}^2-\gamma_{\mu}^2}$
taken to be real (the bath oscillators are underdamped), the
formal solutions for Eq.~(\ref{cde:2}) can be written in terms of
the oscillator's initial position $q_{\mu 0}$ and velocity
$\dot{q}_{\mu 0}$ as
\begin{widetext}
\begin{eqnarray}
 q_{\mu}(t)=&& [q_{\mu 0}\cos \Omega_{\mu} t
+\frac{1}{\Omega_{\mu}}(\dot{q}_{\mu 0}+\gamma_{\mu}q_{\mu})\sin
\Omega_{\mu} t]e^{-\gamma_{\mu}t}
+\int_{0}^{t}\frac{1}{m_{\mu}\Omega_{\mu}}\sin \Omega_{\mu}(t-s)
e^{-\gamma_{\mu}(t-s)}F_{\mu}(s)ds \nonumber\\
&&a_{\mu}(x(t))-\frac{1}{\Omega_{\mu}}a_{\mu}(x(0))
[\Omega_{\mu}\cos\Omega_{\mu}t+\gamma_{\mu}\sin\Omega_{\mu}t]
e^{-\gamma_{\mu}t}\nonumber\\
&& -\int_{0}^{t}\frac{1}{m_{\mu}\Omega_{\mu}}
[\Omega_{\mu}\cos\Omega_{\mu}s+\gamma_{\mu}\sin\Omega_{\mu}s]
e^{-\gamma_{\mu}s} \frac{\partial a_{\mu}(x(t-s))}{\partial x}
\dot{x}(t-s)ds.
 \label{cde:4}
\end{eqnarray}
\end{widetext}
Upon substitution of Eq.~(\ref{cde:4}) into Eq.~(\ref{cde:1}) and
rearranging, the equation of motion for the system can be written
\begin{eqnarray}
m \ddot{x} &=& -\frac{\partial U(x)}{\partial x}
-\int_{0}^{t}\eta(x(t),x(s);t,s)\dot{x}(s)ds
\nonumber\\
&&+F_{E1}(t)+F_{E2}(t)+F_{E3}(t)
 \label{cde:5}
\end{eqnarray}
\begin{widetext} where the noise terms are given by
\begin{subequations}
\label{cde:6}
\begin{eqnarray}
F_{E1}(t)&=&\sum_{\mu}m_{\mu}\omega_{\mu}^{2} [q_{\mu
0}-a_{\mu}(x(0))][\cos\Omega_{\mu}t+\frac{\Omega_{\mu}}
{\gamma_{\mu}}\sin\Omega_{\mu}t]e^{-\gamma_{\mu}t},
 \label{cde:6.1}
\end{eqnarray}
\begin{eqnarray}
F_{E2}(t)&=&\sum_{\mu}m_{\mu}\omega_{\mu}^{2} \dot{q}_{\mu
0}[\cos\Omega_{\mu}t+\frac{\sin\Omega_{\mu}t}
{\Omega_{\mu}}]e^{-\gamma_{\mu}t},
 \label{cde:6.2}
\end{eqnarray}
\begin{eqnarray}
F_{E3}(t)&=&\sum_{\mu}\frac{\omega_{\mu}^{2}}{\Omega_{\mu}}
\int_{0}^{t}\sin\Omega_{\mu}(t-s)e^{-\gamma_{\mu}(t-s)}F_{\mu}(s)ds
\frac{\partial a_{\mu}(x(t))}{\partial x}\label{cde:6.3},
\end{eqnarray}
\end{subequations}
\end{widetext}
and the dissipation kernel is given by
\begin{eqnarray}
\eta(x(t),x(s);t,s)= \sum_{\mu}m_{\mu}\omega_{\mu}^{2}\{
[\cos\Omega_{\mu}(t-s)\nonumber\\
+\frac{\Omega_{\mu}} {\gamma_{\mu}}\sin\Omega_{\mu}(t-s)]
e^{-\gamma_{\mu}(t-s)} \frac{\partial a_{\mu}(x(s))}{\partial x}
\frac{\partial a_{\mu}(x(t))}{\partial x}\}\label{cde:eta}.
\end{eqnarray}
In the past, the properties of noise terms like $F_{E1}(t)$ and
$F_{E2}(t)$ have generally been extracted from the statistical
distributions of the initial conditions. With the introduction of
dissipation in the environment, those noise terms can readily be
seen to be transient terms on the time scales of the environment
(as determined by $\gamma_{\mu}$). Thus the details of the initial
state of the environment are not important to the long term system
dynamics.

The correlations $\langle F_{E3}(t)F_{E3}(\tau)\rangle$ of the
remaining noise term depend upon the correlations of the
individual oscillators' noise terms via Eq.~(\ref{cde:efd}), and
can be written
\begin{eqnarray}
\langle F_{E3}(x,t)F_{E3}(x',\tau)\rangle=
k_{B}T\eta(x,x';t,\tau)\nonumber\\
+\frac{1}{4\gamma_{\mu}\omega_{\mu}^{2}}
[\gamma_{\mu}^{2}\cos\Omega_{\mu}(t+\tau)
-\omega_{\mu}^{2}\cos\Omega_{\mu}(t-\tau)\nonumber\\
-\omega_{\mu}\gamma_{\mu}\sin\Omega_{\mu}(t+\tau)]
e^{-\gamma_{\mu}(t+\tau)} \frac{\partial a_{\mu}(x))}{\partial x}
\frac{\partial a_{\mu}(x')}{\partial x'}.
 \label{cde:corr}
\end{eqnarray}
The first term on the right hand side of Eq.~(\ref{cde:corr}) is
the long term correlation function of the effective noise.  The
second term contains a factor of $e^{-\gamma_{\mu}(t+\tau)}$, and
thus is transient for long timescales.

Since $\eta(x,x';t,\tau)$ is a narrow function of $t-\tau$, we can
make a Markov approximation in Eq.~(\ref{cde:5}):
\begin{equation}
m \ddot{x} = -\frac{\partial U(x)}{\partial x}
-\bar{\eta}(x,x)\dot{x}+F_{s}(x,t),
 \label{cde:mkv}
\end{equation}
where
\begin{equation}
\bar{\eta}(x,x')=\sum_{\mu} 2 m_{\mu}\gamma_{\mu} \frac{\partial
a_{\mu}(x))}{\partial x} \frac{\partial a_{\mu}(x')}{\partial x'}.
\label{cde:etabar}
\end{equation}
The corresponding fluctuation-dissipation relation is
\begin{equation}
\langle F_{s}(x,t)F_{s}(x',\tau)\rangle=
2k_{B}T\eta(x,x')\delta(t-\tau).
 \label{cde:fd}
\end{equation}
The role of the spatial correlations of the noise in quantum
decoherence and the lack of importance of those correlations in
classical phenomena has been discussed
elsewhere.\cite{mrglocaldis,mrglocaldis2}

Thus, adding ``fast" thermal relaxation to the environment of the
system of interest lead to Markovian equations of motion with the
usual fluctuation-dissipation relations.  Memory effects due to
the details of the initial environment state (including possible
correlations with the initial system state) are erased on the
environment relaxation time scales.  This result provides
excellent motivation for exploring similar models in the quantum
mechanical domain.

\section{\label{env_prop} Properties of quantum open environment}
As with the classical model, in order to account for the open
nature of the environment we need to incorporate modifications to
the dynamics of the environment degrees of freedom.  The
environment is modelled as a set of independent oscillators whose
evolution is governed by a Markovian master equation:
\begin{equation}
 \frac{\partial \rho}{\partial t}=L[\rho].
 \label{ep:me}
\end{equation}
The generator $L$ of the evolution is taken to be of the Lindblad
form\cite{Lindblad}:
\begin{eqnarray}
 \frac{\partial \rho}{\partial t}&=&L[\rho ] \nonumber \\
 &=&\frac{1}{ i\hbar}[H,\rho ]+\frac{1}{
2\hbar}\sum_{\mu } [{V}_{\mu }\rho ,{V}_{\mu }^{\dagger
}]+[{V}_{\mu },\rho {V}_{\mu }^{\dagger }].
 \label{ep:melf}
\end{eqnarray}

Formally, the solution to Eq.~(\ref{ep:melf}) is
\begin{equation}
\rho(t)=\Lambda(t)[\rho]
 =e^{Lt}[\rho],
\end{equation}
expectation values are defined via the trace operation, allowing
the definition of the adjoint representation of the evolution
operator $\Lambda^{\ast}(t)[O]$ via
\begin{equation}
\mathrm{Tr}[\rho\Lambda^{\ast}(t)[O]]=\mathrm{Tr}[\Lambda(t)[\rho]O],
\end{equation}
and for the generator $L^{\ast}$
\begin{equation}
\mathrm{Tr}[\rho L^{\ast}[O]]=\mathrm{Tr}[L[\rho]O].
\end{equation}
The adjoint representation of $L$ (i.e. Heisenberg picture) is
given by
\begin{equation}
L^{*}[O] =-\frac{1 }{i\hbar}[H,O] +\frac{1 }{ 2\hbar}\sum_{\mu }
{V}_{\mu }^{\dagger }[O,{V}_{\mu }]+[{V}_{\mu }^{\dagger
},O]{V}_{\mu }.
\end{equation}

To model the relaxation of the environment, we will use a subset
of a family of master equations that have been studied extensively
in the literature \cite{sns,Isar+ssss,Isar1,Isar_sns1} where the
Lindblad form of the master equation can be rewritten as:
\begin{eqnarray}
 \frac{\partial \rho}{\partial t}
 &=&\frac{1}{ i\hbar}[H_{0},\rho
 ]-\frac{i}{2\hbar}(\lambda+\mu)[q,\{p,\rho\}]
  \nonumber \\
&&+\frac{i}{2\hbar}(\lambda-\mu)[p,\{q,\rho\}] -
\frac{D_{pp}}{\hbar^{2}}[q,[q,\rho]]
 \nonumber \\
&&-\frac{D_{qq}}{\hbar^{2}}[p,[p,\rho]]
 \nonumber \\
 &&+\frac{D_{pq}}{\hbar^{2}}([q,[p,\rho]] +[p,[q,\rho]]).
 \label{ep:mesns}
\end{eqnarray}
The details of the model are determined by the specification of
the diffusion coefficients $D_{qq}$, $D_{pp}$ and $D_{pq}$ and
damping constants $\mu$ and $\lambda$, subject to the
constraints:\cite{sns}
\begin{eqnarray}
 D_{qq}&>&0
  \nonumber \\
D_{pp}&>&0
  \nonumber \\
D_{qq}D_{pp}-D_{pq}^{2}&\geq&(\frac{\lambda\hbar}{2})^{2}.
 \label{ep:constraint}
\end{eqnarray}
The nominal hamiltonian is
\begin{equation}
H_{0}=\frac{p^{2}}{2m}+\frac{m \omega_{0}^{2}}{2}q^{2}
\end{equation}

$\tilde{\rho}$ is taken to be the (unique) asymptotic state of
$\Lambda(t)$, that is
\begin{equation}
\Lambda_{\infty}[\rho]\equiv\lim_{t\rightarrow\infty}\Lambda(t)[\rho]=\tilde{\rho}
\end{equation}
for any initial environment state $\rho$.  Since $\tilde{\rho}$ is
a stationary state of $L$
\begin{equation}
L[\tilde{\rho}]=0.
\end{equation}
The asymptotic behavior of the environment is encapsulated in
$\Lambda_{\infty}$, so that
\begin{equation}
\Lambda_{\infty}[\rho]=\tilde{\rho}^{(E)}
\end{equation}
for any state $\rho$.  Although $\Lambda_{\infty}$ is defined only
on the space of density operators (including pure states), we will
need to extend its domain. If $\{|e_{k}\rangle\}$ is a basis for
the environment Hilbert space, then
\begin{equation}
\Lambda_{\infty}[|e_{k}\rangle\langle e_{k}|]=\tilde{\rho}
\end{equation}
for any $k$.  Furthermore, if
\begin{eqnarray}
|\psi_{nm}\rangle \equiv\frac{1}{\sqrt{2}}(|e_{n}\rangle
+|e_{m}\rangle)
\nonumber \\
|\chi_{nm}\rangle \equiv\frac{1}{\sqrt{2}}(|e_{n}\rangle
+i|e_{m}\rangle )
\end{eqnarray}
then
\begin{eqnarray}
\Lambda_{\infty}[|\psi_{nm}\rangle \langle
\psi_{nm}|]&=&\tilde{\rho}
\nonumber \\
&=&\frac{1}{2}(\Lambda_{\infty}[|e_{n}\rangle\langle e_{n}|]
+\Lambda_{\infty}[|e_{m}\rangle\langle e_{m}|]
\nonumber \\
&&+\Lambda_{\infty}[|e_{n}\rangle\langle e_{m}|]
+\Lambda_{\infty}[|e_{m}\rangle\langle e_{n}|])
\nonumber \\
&=&\tilde{\rho} +\frac{1}{2}(\Lambda_{\infty}[|e_{n}\rangle\langle
e_{m}|]
\nonumber \\
&&+\Lambda_{\infty}[|e_{m}\rangle\langle e_{n}|]),
\end{eqnarray}
and
\begin{eqnarray}
\Lambda_{\infty}[|\chi_{nm}\rangle \langle
\chi_{nm}|]&=&\tilde{\rho}
\nonumber \\
&=&\frac{1}{2}(\Lambda_{\infty}[|e_{n}\rangle\langle e_{n}|]
+\Lambda_{\infty}[|e_{m}\rangle\langle e_{m}|]
\nonumber \\
&&+i\Lambda_{\infty}[|e_{n}\rangle\langle e_{m}|]
-i\Lambda_{\infty}[|e_{m}\rangle\langle e_{n}|])
\nonumber \\
&=&\tilde{\rho}
+i\frac{1}{2}(\Lambda_{\infty}[|e_{n}\rangle\langle e_{m}|]
\nonumber \\
&&-i\Lambda_{\infty}[|e_{m}\rangle\langle e_{n}|]),
\end{eqnarray}
Which implies
\begin{equation}
\Lambda_{\infty}[|e_{n}\rangle\langle
e_{m}|]=\delta_{nm}\tilde{\rho}.
\end{equation}
For an arbitrary operator $O$
\begin{eqnarray}
\Lambda_{\infty}[O]&=&\sum_{nm}O_{nm}\Lambda_{\infty}[|e_{n}\rangle\langle
e_{m}|]
\nonumber \\
&=&\sum_{m}O_{m m}\tilde{\rho}
\nonumber \\
&=&\mathrm{Tr}[O]\tilde{\rho} \label{ep:tracer}
\end{eqnarray}

 There is in general no guarantee that an asymptotic
state exists for arbitrary $L$, that is, not all choices for the
parameters $D_{qq}$, $D_{pp}$, etc. will be appropriate to model
the a system dynamically relaxing to equilibrium.  For example, if
the parameters satisfy
\begin{subequations}
\label{ep:gibbs}
\begin{equation}
D_{pp}=\frac{\lambda+\mu}{2}\hbar m
\omega\coth{\frac{\hbar\omega}{2k_{B}T}},
\end{equation}
\begin{equation}
D_{qq}=\frac{\lambda-\mu}{2}\frac{\hbar}{m\omega}\coth{\frac{\hbar\omega}{2k_{B}T}},
\end{equation}
\begin{equation}
D_{pq}=0,
\end{equation}
\end{subequations}
then the Gibbs state is the asymptotic state.\cite{sns,Isar+ssss}
On the other hand, if
\begin{subequations}
\label{ep:wierd}
\begin{equation}
D_{qq}=\frac{\hbar \lambda }{2m\Omega},
\end{equation}
\begin{equation}
D_{pp}=\frac{\hbar \lambda m \omega^{2}}{2\Omega},
\end{equation}
\begin{equation}
D_{pq}=-\frac{\hbar \lambda \mu}{2\Omega},
\end{equation}
\end{subequations}
(where $\Omega^2=\omega^2-\mu^2$) there can be persistent pure
states.\cite{Isar1,Isar_sns1} Since we are considering the primary
effect of the openness of the environment to be effectiveness
which relax the environment towards an equilibrium state, we would
only consider those choices of parameters for which there is a
unique asymptotic state.

Sandulescu and Scutaru have determined the time dependence and
asymptotic behavior of various moments of $p$ and $q$, which will
be useful in later calculations.  The evolution of the first order
moments is given by
\begin{subequations}
\label{ep:momt1}
\begin{equation}
\frac{\partial \langle q\rangle}{\partial t}
=-(\lambda-\mu)\langle q\rangle+\frac{1}{m}\langle p\rangle
\end{equation}
\begin{equation}
\frac{\partial \langle p\rangle}{\partial t}
=-(\lambda+\mu)\langle p\rangle-m\omega^{2}\langle p\rangle,
\end{equation}
\end{subequations}
which illustrates the role of the dissipation coefficients
$\lambda$ and $\mu$. With
\begin{equation}
S_{pq}\equiv \frac{1}{2}\{p,q\},
\end{equation}
the evolution of the second order moments is given by
\begin{subequations}
\label{ep:momt2}
\begin{equation}
\frac{\partial \langle p^{2}\rangle}{\partial t}
=-2m\omega^{2}\langle S_{pq} \rangle-2(\lambda+\mu)\langle
p^{2}\rangle+2D{pp}
\end{equation}
\begin{equation}
\frac{\partial \langle q^{2}\rangle}{\partial t}
=\frac{2}{m}\langle S_{pq} \rangle-2(\lambda-\mu)\langle
q^{2}\rangle+2D{qq}
\end{equation}
\begin{equation}
\frac{\partial \langle S_{pq}\rangle}{\partial t}
=\frac{1}{m}\langle p^{2}\rangle-m\omega^{2}\langle q^{2}
\rangle-2\lambda\langle S_{pq}\rangle+2D{pq}
\end{equation}
\end{subequations}
which illustrates the role of the diffusion coefficients $D_{pp}$,
$D_{qq}$ and $D_{pq}$.

The asymptotic first order moments are
\begin{subequations}
\label{ep:moma}
\begin{equation}
\langle p\rangle_{\infty} = 0 \label{ep:moma1}
\end{equation}
\begin{equation}
\langle q\rangle_{\infty} = 0 \label{ep:moma2}.
\end{equation}
\end{subequations}
For the second order moments we have
\begin{eqnarray}
\langle p^2\rangle_{\infty} &=& \frac{1}{2\lambda(\lambda^2
+\omega^2-\mu^2)}
\{m^{2}\omega^{4}D_{qq}
\nonumber\\
&&+[2\lambda(\lambda-\mu)+\omega^2]D_{pp}
\nonumber\\
&&-2m\omega^2[\lambda-\mu]D_{pq}\},
 \label{ep:mompp}
\end{eqnarray}
\begin{eqnarray}
\langle q^2\rangle_{\infty} &=&
\frac{1}{2(m\omega)^{2}\lambda(\lambda^2+\omega^2-\mu^2)} [
(m\omega)^{2}\omega^{2}D_{qq}
\nonumber\\
&&+\omega^{2} D_{pp}+2m\omega^2(\lambda+\mu)D_{pq}],
 \label{ep:momqq}
\end{eqnarray}
and
\begin{eqnarray}
\langle S_{pq}\rangle_{\infty} &=&\langle
\frac{1}{2}\{p,q\}\rangle_{\infty}
\nonumber\\
&=&\frac{1}{2(m\lambda(\lambda^2+\omega^2-\mu^2)} [
-(\lambda+\mu)(m\omega)^{2}D_{qq}
\nonumber\\
&&+(\lambda-\mu)D_{pp}+2m(\lambda^{2}-\mu^{2})D_{pq}].
 \label{ep:mompq}
\end{eqnarray}
We will also need the time dependence of $q(t)$ in the Heisenberg
picture.  This can be readily extracted from Sandulescu and
Scutaru 's results for the time dependence for the first order
moments.\cite{sns} The result is
\begin{eqnarray}
q_{H}(t)&=&\Lambda(t)^{\ast}[q]\nonumber\\
&=&q(\cos{\Omega t}+\frac{\mu}{\Omega}\sin{\Omega t})e^{-\lambda
t} \nonumber\\
&+&\frac{p}{m\Omega}\sin{\Omega t}\, e^{-\lambda t},
 \label{ep:momqt}
\end{eqnarray}
with
\begin{equation}
\Omega = \sqrt{\omega^2-\mu^2}
\end{equation}
and $\Omega$ taken to be real (the oscillators are underdamped).

The specific environment we will employ consists of a set of
independent oscillators, each subject to evolution of the form
Eq.~(\ref{ep:mesns}), with position operators $\{q_{i}\}$ and
associated parameters $\{\mu_{i}\}$, $\{\lambda_{i}\}$, etc.  Thus
the Hilbert space of the environment is actually the tensor
product of the Hilbert space corresponding to each environment
degree of freedom.  The interaction between the environment and
system is accounted for by adding an interaction term to the
composite system's hamiltonian:
\begin{equation}
U_{I}=\sum_{n} V_{n}^{(S)}\otimes q_{n}^{(E)}.
 \label{ep:interaction}
\end{equation}
We will need the correlation functions $\langle
q_{n}(t)q_{m}\rangle$ in the next section.  Using
Eq.~(\ref{ep:momqt}), these correlations can be written in terms
of the asymptotic correlations as
\begin{eqnarray}
\langle
q_{n}(t)q_{m}\rangle&=&\mathrm{Tr}[\Lambda^{\ast}(t)[q_{n}]\,
q_{m}\,\tilde{\rho}]
\nonumber\\
&=&\delta_{nm}[\langle q_{n}^{2}\rangle_{\infty}(\cos{\Omega
t}+\frac{\mu}{\Omega}\sin{\Omega t})e^{-\lambda
t} \nonumber\\
&+&\frac{2\langle S_{pqn}\rangle_{\infty}-i\hbar}{2
m\Omega}\sin{\Omega t}\, e^{-\lambda t}],
 \label{ep:corr_ftn}
\end{eqnarray}
where we have made use of the relation
\begin{eqnarray}
p_{n}q_{n}&=&\frac{1}{2}(\{p_{n},q_{n}\}+[p_{n},q_{n}])
\nonumber\\
&=&S_{pqn}-\frac{i\hbar}{2}.
\end{eqnarray}

\section{\label{proj} Quantum master equation from an open environment}

To construct the master equation, we will use the projection
operator method in the weak coupling regime, largely following the
presentation of Alicki and Lendi.\cite{Alicki+Lendi} The composite
system (the system of interest plus its environment) is taken to
evolve according to Eq.~(\ref{ep:me}).  The generator of this
evolution is the combination of the nominal dynamics and an
interaction:
\begin{equation}
L=L_{0}+L_{I},
\end{equation}
where the system and environment dynamics each contribute
separately:
\begin{equation}
L_{0}=L_{S}+L_{E}. \label{qos:Ladd}
\end{equation}
The generators for the system and environment nominal evolutions
act on on the corresponding subspaces, so that
\begin{equation}
L_{S}[A^{(S)}\otimes B^{(E)}]=L_{S}^{(S)}[A^{(S)}]\otimes B^{(E)},
\end{equation}
and
\begin{equation}
L_{E}[A^{(S)}\otimes B^{(E)}]=A^{(S)}\otimes L_{E}^{(E)}[B^{(E)}]
\end{equation}
for all $A^{(S)}$ and $B^{(E)}$. $L_{S}^{(S)}$ is taken to be
unitary, and $L_{E}^{(E)}$ is taken to be of the Lindblad form. It
is useful to note that necessarily $L_{S}$ and $L_{E}$ commute.

The projection operator is defined in terms of a partial trace:
\begin{equation}
P_{0}[O]=\mathrm{Tr}_{E}[O]\otimes \tilde{\rho}^{(E)}.
\end{equation}
We wish to establish some important relations between $P_{0}$ and
the generators $L_{S}$ and $L_{E}$.
\begin{eqnarray}
L_{E}P_{0}[O]&=&L_{E}[\mathrm{Tr}_{E}[O]\otimes
\tilde{\rho}^{(E)}]
 \nonumber \\
 &=&\mathrm{Tr}_{E}[O]\otimes
L_{E}^{(E)}[\tilde{\rho}^{(E)}]
 =0
  \label{qos:pr1}
\end{eqnarray}
 for any operator $O$ so that $L_{E}P_{0}=0$. $L_{E}^{(E)}$ generates trace preserving evolution
so that
\begin{equation}
\mathrm{Tr}_{E}[L_{E}^{(E)}[O^{(E)}]]=0.
\end{equation}
for all $O^{(E)}$. Thus
\begin{eqnarray}
P_{0}L_{E}[A^{(S)}\otimes B^{(E)}] &=&P_{0}[A\otimes
L_{E}^{(E)}[B^{(E)}]
\nonumber \\
&=&(\mathrm{Tr}_{E}[L_{E}^{(E)}[B^{(E)}]])
A^{(S)}\otimes\tilde{\rho}^{(E)}
\nonumber \\
 &=&0,
  \label{qos:pr2}
\end{eqnarray}
for all $A^{(S)}$ and $B^{(E)}$ so that in general $P_{0}L_{E}=0$.
\begin{eqnarray}
P_{0}L_{S}[A^{(S)}\otimes B^{(E)}]
&=&P_{0}[L_{S}^{(S)}[A^{(S)}]\otimes B^{(E)}
\nonumber \\
&=&L_{S}^{(S)}[A^{(S)}]\otimes \tilde{\rho}^{(E)}
\mathrm{Tr}^{(E)}[B^{(E)}]
\nonumber \\
 &=&L_{S}[[A^{(S)}]\otimes \tilde{\rho}^{(E)}
\mathrm{Tr}^{(E)}[B^{(E)}]]\nonumber \\
&=&L_{S}P_{0}[A^{(S)}\otimes B^{(E)}]
  \label{qos:pr3}
\end{eqnarray}
for all $A^{(S)}$ and $B^{(E)}$ so that in general
$P_{0}L_{S}=L_{S}P_{0}$.  Defining a second projection
$P_{1}\equiv1-P_{0}$, it is easy to see that
$P_{1}L_{S}=L_{S}P_{1}$.  In order to focus on the reduced
dynamics of the system of interest, we study the dynamics of
$P_{0}\rho$ in certain approximations. Applying the projectors to
the master equation for the composite system and making use of the
idempotent property of projections produces the following
equations:
\begin{subequations}
\begin{equation}
 \frac{\partial P_{0}\rho}{\partial
 t}=P_{0}LP_{0}P_{0}\rho+P_{0}LP_{1}P_{1}\rho,
 \label{qos:eom1}
\end{equation}
\begin{equation}
 \frac{\partial P_{1}\rho}{\partial
 t}=P_{1}LP_{1}P_{1}\rho+P_{1}LP_{0}P_{0}\rho.
 \label{qos:eom2}
\end{equation}
\end{subequations}
 Eq.~(\ref{qos:eom2}) can be formally integrated and substituted
 into  Eq.~(\ref{qos:eom1}), so that
\begin{eqnarray}
\frac{\partial P_{0}\rho}{\partial t}&=&P_{0}L P_{0}P_{0}\rho(t)
\nonumber \\&&+P_{0}L P_{1}[ e^{P_{1}L P_{1}t}P_{1}\rho(0)
\nonumber \\
&&+\int_{0}^{t} e^{P_{1}L P_{1}(t-s)}P_{1}L P_{0}P_{0}\rho(s)]ds.
\label{qos:eom3}
\end{eqnarray}
Using the relations between the projectors and generators
discussed above, we have
\begin{eqnarray}
P_{0}L P_{1}&=&P_{0}L_{I}P_{1},
\nonumber \\
P_{1}L P_{0}&=&P_{1}L_{I}P_{0},
\nonumber \\
P_{0}L P_{0}&=&L_{S}P_{0}+P_{0}L_{I}P_{0},
\label{qos:pr4}
\end{eqnarray}
so that  Eq.~(\ref{qos:eom3}) can be written
\begin{eqnarray}
\frac{\partial P_{0}\rho}{\partial
t}&=&(L_{S}+P_{0}L_{I}P_{0})P_{0}\rho(t)+P_{0}L_{I}
P_{1}e^{Lt}P_{1}\rho(0)
\nonumber \\
&&+\int_{0}^{t} P_{0}L_{I} P_{1}e^{L(t-s)}P_{1}L_{I}
P_{0}P_{0}\rho(s) ds.
\label{qos:eom4}
\end{eqnarray}
This result is exact, within the constraints placed on the model
so far.

The interaction, as specified by  Eq.~(\ref{ep:interaction}), can
be written
\begin{eqnarray}
L_{I}[O]&=&\frac{1}{i\hbar}[U_{I},O]
\nonumber \\
&=&\frac{1}{i\hbar}\sum_{n} [V_{n}^{(S)}\otimes q_{n}^{(E)},O].
\label{qos:interaction}
\end{eqnarray}
This form, along with the first order moments of the environment
asymptotic state, simplifies the first term of the right hand side
of Eq.~(\ref{qos:eom4}). Specifically, for any operator $O$,
\begin{widetext}
\begin{eqnarray}
P_{0}L_{I}P_{0}[O]&=&\mathrm{Tr}_{E}[\frac{1}{i\hbar}\sum_{n}
[V_{n}^{(S)}\otimes q_{n}^{(E)},
\mathrm{Tr}_{E}[O]\otimes\tilde{\rho}^{(E)}]]
\otimes\tilde{\rho}^{(E)}
\nonumber \\
&=&\frac{1}{i\hbar}\sum_{n}\mathrm{Tr}^{(E)}
[q_{n}^{(E)}\tilde{\rho}^{(E)}]
[V_{n}^{(S)},\mathrm{Tr}_{E}[O]]\otimes\tilde{\rho}^{(E)}
=\frac{1}{i\hbar}\sum_{n}\langle q_{n}^{(E)}\rangle_{\infty}
[V_{n}^{(S)},\mathrm{Tr}_{E}[O]]\otimes\tilde{\rho}^{(E)} .
  \label{qos:termLi}
\end{eqnarray}
\end{widetext}
From Eq.~(\ref{ep:moma}) each term in the sum has a factor
$\langle q_{n}^{(E)}\rangle_{\infty}=0$ so that
\begin{equation}
P_{0}L_{I}P_{0}=0.
 \label{qos:plip}
\end{equation}
Furthermore
\begin{equation}
P_{0}L_{I}P_{1}=P_{0}L_{I}(1-P_{0})=P_{0}L_{I}
 \label{qos:plip1}
\end{equation}
and
\begin{equation}
P_{1}L_{I}P_{0}=(1-P_{0})L_{I}P_{0}=L_{I}P_{0},
 \label{qos:p1lip}
\end{equation}
so that Eq.~(\ref{qos:eom4}) can be written
\begin{eqnarray}
\frac{\partial P_{0}\rho}{\partial t}&=& L_{S}P_{0}\rho(t)
\nonumber \\
&+&P_{0}L_{I} P_{1}e^{Lt}P_{1}\rho(0)
\nonumber \\
&+&\int_{0}^{t} P_{0}L_{I} e^{L(t-s)}L_{I} P_{0}P_{0}\rho(s) ds].
\label{qos:eom5}
\end{eqnarray}

The second term on the right hand side of Eq.~(\ref{qos:eom5})
\begin{equation}
P_{0}L_{I} P_{1}e^{Lt}P_{1}\rho(0)
\label{qos:transient}
\end{equation}
represents a transient term which depends upon the initial state
of the composite system.  For most constructive models the initial
state is taken to be a factored state of an arbitrary system state
and the environment in its asymptotic state:
\begin{equation}
\rho(0)=\rho^{(S)}(0)\otimes\tilde{\rho}^{(E)}.
\end{equation}
For this type of initial condition $\rho(0)=P_{0}\rho(0)$ so that
\begin{equation}
P_{1}\rho(0)=0.
\end{equation}
However, because we have added relaxation to the dynamics of the
environment, this type of factoring assumption is not necessary.
We will be taking the weak coupling limit and so we will show that
Eq.~(\ref{qos:transient}) is (approximately) zero to the lowest
nonvanishing order of the interaction in the remaining terms of
Eq.~(\ref{qos:eom4}),  which turns out to be second order.
Formally we can write
\begin{equation}
e^{Lt} = e^{L_0 t}+\int_{0}^{t}e^{L(t-s)}L_{I}e^{L_0 s}ds.
\label{qos:intrep}
\end{equation}
Substituting Eq.~(\ref{qos:intrep}) into Eq.~(\ref{qos:transient})
and keeping only second order, we have the approximation
\begin{eqnarray}
P_{0}L_{I} P_{1}e^{Lt}P_{1}\rho(0)\approx P_{0}L_{I} P_{1}e^{L_{0}
t}P_{1}\rho(0)
\nonumber \\
+P_{0}L_{I} P_{1}\int_{0}^{t}e^{L_0(t-s)}L_{I}e^{L_{0} s}ds
P_{1}\rho(0). \label{qos:trans2}
\end{eqnarray}
We are interested in timescales $t$ which are assumed to be much
longer than the relaxation timescales of the environment, so that
\begin{eqnarray}
e^{L_{0}t}&=&e^{L_{S}t}e^{L_{E}t}
\nonumber \\
&\approx& e^{L_{S}t}\Lambda_{E\infty}. \label{qos:transfactor}
\end{eqnarray}
Similarly, for the integral from $t$ to $s$, either $t$ or $t-s$
(or both) is large compared to the relaxation timescales of the
environment, so that either
\begin{equation}
e^{L_{0}(t-s)}\approx e^{L_{S}(t-s)}\Lambda_{E\infty}
\label{qos:trpr1}
\end{equation}
or
\begin{equation}
e^{L_{0}s}\approx e^{L_{S}s}\Lambda_{E\infty} \label{qos:trpr2}
\end{equation}
or both.  From Eq.~(\ref{ep:tracer}), for an operator $O$
decomposed in terms of the basis for the environment
$\{|e_{k}\rangle\}$ and the basis for the system
$\{|\phi_{\nu}\rangle\}$ we have
\begin{eqnarray}
\Lambda_{E\infty}[O]&=&O_{\alpha m ,\beta n
}\Lambda_{E\infty}[|\phi_{\alpha}\rangle\otimes|e_{m}\rangle
\langle \phi_{\alpha}|\otimes\langle e_{m}|]
\nonumber \\
&=&O_{\alpha m ,\beta n }|\phi_{\alpha}\rangle\langle
\phi_{\alpha}|\otimes\Lambda_{E\infty}^{(E)}[|e_{m}\rangle \langle
e_{m}|]
\nonumber \\
&=&O_{\alpha m ,\beta n }|\phi_{\alpha}\rangle\langle
\phi_{\alpha}|\otimes \delta_{mn}\tilde{\rho}^{(E)}
\nonumber \\
&=&\mathrm{Tr}_{E}[O]\otimes \tilde{\rho}^{(E)}=P_{0}O.
\label{qos:prj}
\end{eqnarray}
When Eq.~(\ref{qos:prj}) is applied to Eq.~(\ref{qos:trans2}) with
either Eq.~(\ref{qos:prj}) or Eq.~(\ref{qos:prj}) applying, then
all terms in Eq.~(\ref{qos:trans2}) end up with factors of either
$P_{0}P_{1}$ or $P_{1}P_{0}$, both of which are $0$.  Using this
result and keeping only up to second order in the interaction
term, the master equation Eq.~(\ref{qos:eom5}) becomes
\begin{eqnarray}
\frac{\partial P_{0}\rho}{\partial t}&=&L_{S}P_{0}\rho(t)
\nonumber \\
&&+\int_{0}^{t} P_{0}L_{I} e^{L_{0}(t-s)}L_{I} P_{0}P_{0}\rho(s)
ds. \label{qos:eom6}
\end{eqnarray}

We can rewrite  Eq.~(\ref{qos:eom6})using Eq.~(\ref{qos:plip1})
and Eq.~(\ref{qos:p1lip}) to get
\begin{eqnarray}
\frac{\partial P_{0}\rho}{\partial t}&=&L_{S}P_{0}\rho(t)
\nonumber \\
&+&\int_{0}^{t} P_{0}L_{I}P_{1} e^{L_{0}(t-s)}P_{1}L_{I}
P_{0}P_{0}\rho(s) ds. \label{qos:eomt1}
\end{eqnarray}
In the integrand in Eq.~(\ref{qos:eomt1}) we see the factor
$P_{1}e^{L_{0}(t-s)}P_{1}$ which (following the discussion above)
will be zero for $t-s$ longer than environment relaxation
timescales so that the primary contribution to the integral is for
$s\approx t$. This yields the simplest Markovian master equation
we will extract from the model:
\begin{eqnarray}
\frac{\partial P_{0}\rho}{\partial t}&=&L_{S}P_{0}\rho(t)
\nonumber \\
&&+ P_{0}L_{I}L_{I} P_{0}\rho(t). \label{qos:eomt2}
\end{eqnarray}
Using the particular form of the interaction given by
Eq.~(\ref{qos:interaction}) and noting that the independence of
the oscillators comprising the environment implies
\begin{equation}
\langle q_{n}q_{m}\rangle_{\infty}=\delta_{nm}\langle
q_{n}^{2}\rangle_{\infty}
\end{equation}
allows us to rewrite Eq.~(\ref{qos:interaction}) in a final form:
\begin{eqnarray}
\frac{\partial P_{0}\rho}{\partial t}&=&L_{S}P_{0}\rho(t)
\nonumber \\
&&-\frac{1}{\hbar^2} \sum_{n}\langle
q_{n}^{2}\rangle_{\infty}[V_{n},[V_{n}, P_{0}\rho(t)]].
\label{qos:eomt3}
\end{eqnarray}
Eq.~(\ref{qos:eomt3}) represents the added effect of environment
induced noise on the system's dynamics which is responsible for
phenomena such as quantum decoherence.  However, additional
effects such as dissipation are not present and will require a
more careful handling of the Markov approximations.

To reconsider the Markov approximations, we return  to
Eq.~(\ref{qos:eom6}). The unitarity of the isolated system's
evolution implies that
\begin{equation}
e^{L_{S}t}[A B]=e^{L_{S}t}[A]e^{L_{S}t}[B].
\label{qos:unitarydist}
\end{equation}
The naive Markov approximation is introduced into
Eq.~(\ref{qos:eom6}) by examining
\begin{eqnarray}
 e^{L_{0}(t-s)}&L_{I}& P_{0}\rho(s) =
 e^{L_{E}(t-s)}e^{L_{S}(t-s)}L_{I} P_{0}\rho(s)
\nonumber \\
&=&\frac{1}{i\hbar}
e^{L_{E}(t-s)}e^{L_{S}(t-s)}[U_{I},P_{0}\rho(s)]
\nonumber \\
&=&\frac{1}{i\hbar}
e^{L_{E}(t-s)}[(e^{L_{S}(t-s)}U_{I}),(e^{L_{S}(t-s)}P_{0}\rho(s))]
\nonumber \\
&\approx&\frac{1}{i\hbar}
e^{L_{E}(t-s)}[(e^{L_{S}(t-s)}U_{I}),P_{0}\rho(t)]
\label{qos:nmap}
\end{eqnarray}
The Markovian master equation can be written as
\begin{equation}
\frac{\partial P_{0}\rho}{\partial t}=L_{S}P_{0}\rho(t) + K
P_{0}\rho(t)
\end{equation}
where
\begin{equation}
K[P_{0}\rho]=\int_{0}^{\infty}P_{0}L_{I}e^{L_{0}s}L_{I}P_{0}\,
P_{0}\rho \, ds,
 \label{qos:defK}
\end{equation}
\begin{widetext}
which becomes for our particular model:
\begin{eqnarray}
K[P_{0}\rho(t)]&=&-\frac{1}{\hbar^{2}}\int_{0}^{\infty}
P_{0}[U_{I},e^{L_{E}(s)}[(e^{L_{S}(s)}U_{I}),P_{0}\rho(t)]ds
\nonumber \\
&=&-\frac{1}{\hbar^{2}}\int_{0}^{\infty}ds\{\sum_{nm}
\mathrm{Tr}_{E}[[V_{n}^{(S)}\otimes q_{n}^{(E)},
e^{L_{E}(s)}[(e^{L_{S}(s)}(V_{m}^{(S)}\otimes q_{m}^{(E)})),
P_{0}\rho(t)]]]\otimes \tilde{\rho}^{(E)} \}. \label{qos:eomk1}
\end{eqnarray}
\end{widetext}
Since $L_{E}$ acts only on the environment, we can make use of the
cyclic property of the trace over the environment to rewrite
Eq.~(\ref{qos:eomk1}) as
\begin{eqnarray}
K[P_{0}\rho(t)]&=&
-\frac{1}{\hbar^{2}}\int_{0}^{\infty}ds\{\sum_{nm}
\mathrm{Tr}_{E}[[V_{n}^{(S)}\otimes q_{n}^{(E)}(s),
\nonumber \\
&&[V_{m}^{(S)}(s)\otimes q_{m}^{(E)}, P_{0}\rho(t)]]]\otimes
\tilde{\rho}^{(E)} \},\label{qos:eomk3}
\end{eqnarray}
where
\begin{subequations}
\label{qos:optime}
\begin{equation}
V_{m}^{(S)}(s)\equiv e^{L_{S}^{(S)}(s)}V_{m}^{(S)},
\label{qos:optime1}
\end{equation}
\begin{equation}
q_{n}^{(E)}(s)\equiv \Lambda_{E}^{(E) \ast}(s)q_{n}^{(E)}.
\label{qos:optime2}
\end{equation}
\end{subequations}
In terms of the correlation functions specified by
Eq.~(\ref{ep:corr_ftn}) we can write
\begin{eqnarray}
K[P_{0}\rho(t)]=&& -\frac{1}{\hbar^{2}}\int_{0}^{\infty}ds
\sum_{n}\{
\nonumber \\
&&\langle q_{n}(s)q_{n}\rangle V_{n}V_{n}(s) P_{0}\rho (t)
\nonumber \\
&-&\langle q_{n}q_{n}(s)\rangle V_{n}P_{0}\rho (t)V_{n}(s)
\nonumber \\
&-&\langle q_{n}(s)q_{n}\rangle V_{n}(s)P_{0}\rho (t)V_{n}
\nonumber \\
&+&\langle q_{n}q_{n}(s)\rangle P_{0}\rho (t)V_{n}(s)V_{n}
\},\label{qos:eomk4}
\end{eqnarray}
where $V_{n}\equiv V_{n}^{(S)}\otimes \mathbb{I}^{(E)}$ and $
\mathbb{I}^{(E)}$ is simply the identify operator on the
environment subspace.  We note that here $V_{n}(s)=e^{L_{S}
s}V_{n}$ is {\it not} the Heisenberg evolved operator, that is
$V_{n}(s)\neq e^{L^{\ast}_{S} s}V_{n}$.

 While Eq.~(\ref{qos:eomk3}) is similar to previous results,
the main difference comes from the nature of  $L_{E}^{(E)}(t)$.
Previous authors have taken $L_{E}^{(E)}(t)$ to necessarily
generate unitary evolution in order to use properties analogous to
Eq.~(\ref{qos:unitarydist}) which does not generally apply to
nonunitary evolution.

Since Eq.~(\ref{qos:eomk3}) has the same mathematical structure as
results obtained without dissipative effects in the environment,
we expect similar shortcomings.  In particular, to insure complete
positivity for the reduced dynamics, we will now apply an
averaging process (sometimes refereed to as the Rotating Wave
Approximation),\cite{Davies, Alicki+Lendi} defined by:
\begin{equation}
\overline{K}=\lim_{a\rightarrow\infty}\frac{1}{a}\int_{0}^{a}e^{-L_{0}
\tau }Ke^{L_{0} \tau } d \tau .
 \label{qos:avgK}
\end{equation}
Since $L_{E}P_{0}$
\begin{equation}
e^{L_{E} \tau }P_{0}=P_{0},
 \label{qos:for_env_avg}
\end{equation}
so that with Eq.~(\ref{qos:Ladd}) and Eq.~(\ref{qos:defK}) we can
write
\begin{equation}
\overline{K}=\lim_{a\rightarrow\infty}\frac{1}{a}\int_{0}^{a}e^{-L_{S}
\tau }K e^{L_{S} \tau } d \tau .
 \label{qos:avgK2}
\end{equation}
Using Eq.~(\ref{qos:unitarydist}) and
Eq.~(\ref{qos:eomk1}),Eq.~(\ref{qos:avgK2}) becomes
\begin{eqnarray}
\overline{K}[P_{0}\rho(t)]&=&-\frac{1}{\hbar^{2}}\lim_{a\rightarrow\infty}\frac{1}{a}
\int_{0}^{a}d\tau \int_{0}^{\infty}ds \sum_{n}\{
\nonumber \\
&&\langle q_{n}(s)q_{n}\rangle V_{n}(-\tau)V_{n}(s-\tau) P_{0}\rho
(t)
\nonumber \\
&-&\langle q_{n}q_{n}(s)\rangle V_{n}(-\tau)P_{0}\rho
(t)V_{n}(s-\tau)
\nonumber \\
&-&\langle q_{n}(s)q_{n}\rangle V_{n}(s-\tau)P_{0}\rho
(t)V_{n}(-\tau)
\nonumber \\
&+&\langle q_{n}q_{n}(s)\rangle P_{0}\rho
(t)V_{n}(s-\tau)V_{n}(-\tau) \}.\; \label{qos:avgK4}
\end{eqnarray}
The operators $\{V_{n}\}$ can be decomposed in terms of the energy
eigenstates of the system hamiltonian $H_{(S)}$:
\begin{eqnarray}
V_{n}&=&\sum_{\mu\nu}
|\mu\rangle\langle\mu |V_{n}|\nu\rangle\langle \nu|\nonumber \\
&=&\sum_{\Delta\omega}\sum_{\epsilon_{\mu}-\epsilon_{\nu}=\hbar\Delta\omega}
|\mu\rangle\langle\mu |V_{n}|\nu\rangle\langle \nu| \nonumber \\
&=&\sum_{\Delta\omega}V_{n,\Delta\omega}. \label{qos:decompV}
\end{eqnarray}
It is clear that with this decomposition
\begin{equation}
V_{n,-\Delta\omega}=V_{n,\Delta\omega}^{\dag}.
 \label{qos:V_adj}
\end{equation}
The time dependence of the operators $\{V_{n}(s)\}$ becomes
\begin{equation}
V_{n}(s)=\sum_{\Delta\omega}e^{-i\Delta\omega
s}V_{n,\Delta\omega}.
 \label{qos:V_t}
\end{equation}
When Eq.~(\ref{qos:V_t}) is substituted into
Eq.~(\ref{qos:avgK4}), there will be oscillating terms which the
integral over $\tau$ will cancel, via
\begin{equation}
\lim_{a\rightarrow\infty}\frac{1}{a} \int_{0}^{a}d\tau
e^{i(\Delta\omega' + \Delta\omega)}=\delta_{\Delta\omega' ,
-\Delta\omega}
 \label{qos:phased}
\end{equation}
so that Eq.~(\ref{qos:avgK4}) becomes
\begin{eqnarray}
\overline{K}[P_{0}\rho(t)]&=&-\frac{1}{\hbar^{2}}\int_{0}^{\infty}ds
\sum_{n,\Delta\omega}e^{-i\Delta\omega \, s}\{
\nonumber \\
&&\langle q_{n}(s)q_{n}\rangle
V_{n,\Delta\omega}V_{n,\Delta\omega}^{\dag} P_{0}\rho (t)
\nonumber \\
&&-\langle q_{n}q_{n}(s)\rangle V_{n,\Delta\omega}P_{0}\rho
(t)V_{n,\Delta\omega}^{\dag}
\nonumber \\
&&-\langle q_{n}(s)q_{n}\rangle V_{n,\Delta\omega}^{\dag}P_{0}\rho
(t)V_{n,\Delta\omega}
\nonumber \\
&&+\langle q_{n}q_{n}(s)\rangle P_{0}\rho
(t)V_{n,\Delta\omega}^{\dag}V_{n,\Delta\omega} \}.\;
\label{qos:avgK5}
\end{eqnarray}
With the definition
\begin{equation}
h_{n,\Delta\omega}+i S_{n,\Delta\omega}\equiv \int_{0}^{\infty}ds
e^{-i\Delta\omega \, s} \langle q_{n}(s)q_{n}\rangle,
\label{qos:hdef}
\end{equation}
and using Eq.~(\ref{qos:V_adj}) we can write Eq.~(\ref{qos:avgK5})
as
\begin{eqnarray}
\overline{K}[P_{0}\rho(t)]&=&
-\frac{i}{\hbar}[\sum_{n,\Delta\omega}
\frac{1}{\hbar}S_{n,\Delta\omega}V_{n,\Delta\omega}
V_{n,\Delta\omega}^{\dag}, P_{0}\rho (t)]
\nonumber \\
&&+\frac{1}{\hbar^{2}}\sum_{n,\Delta\omega}h_{n,\Delta\omega}(
 [V_{n,\Delta\omega}^{\dag}P_{0}\rho
(t),V_{n,\Delta\omega}] \nonumber \\&&+
[V_{n,\Delta\omega}^{\dag},P_{0}\rho (t)V_{n,\Delta\omega}]).
\label{qos:avgK6}
\end{eqnarray}
The first term in the right hand side of Eq.~(\ref{qos:avgK6}) is
simply an additional hamiltonian term.  The remaining terms (those
with $h_{n,\Delta\omega}$) are in the Lindblad form \emph{if}
$h_{n,\Delta\omega}$ is positive. Using Eq.~(\ref{ep:corr_ftn})
and the moments given in Eq.~(\ref{ep:momqq}) and
Eq.~(\ref{ep:mompq}) in Eq.~(\ref{qos:hdef}) we find
\begin{widetext}
\begin{equation}
h_{n,\Delta\omega}=
\frac{[(\lambda_{n}+\mu_{n})^{2}+\Delta\omega^{2}]m_{n}^{2}
D_{qqn}+D_{ppn}+m_{n}\lambda_{n}\hbar\Delta\omega+2(\lambda_{n}+\mu_{n})m_{n}
D_{pqn}}
{[\lambda_{n}^{2}+(\Omega_{n}+\Delta\omega)^2][\lambda_{n}^{2}+(\Omega_{n}-\Delta\omega)^2]}.
\label{qos:h2}
\end{equation}
and
\begin{equation}
S_{n,\Delta\omega}=
\frac{C_{0n}+C_{1n}\Delta\omega+C_{2n}\Delta\omega^{2}+C_{3n}\Delta\omega^{3}}
{2\lambda m^2
[\lambda_{n}^{2}+(\Omega_{n}+\Delta\omega)^2][\lambda_{n}^{2}+(\Omega_{n}-\Delta\omega)^2]}.
\label{qos:s2}
\end{equation}
where
\begin{subequations}
\label{qos:sc}
\begin{equation}
C_{0n}=\hbar m_{n}\lambda_{n}(\lambda_{n}^{2}+\Omega_{n}^{2})^{2},
 \label{qos:sc0}
\end{equation}
\begin{equation}
C_{1n}=[(\Omega_{n}^{2}-3\lambda_{n}^{2})(\mu+\lambda)^2
+(\lambda_{n}^{2}+\Omega_{n}^{2})^{2}]m_{n}^{2}D_{qqn}+(\Omega_{n}^{2}-3\lambda_{n}^{2})
(2(\mu_{n}+\lambda_{n})m_{n}D_{pqn}+D_{ppn}),
 \label{qos:sc1}
\end{equation}
\begin{equation}
C_{2n}=-\hbar \lambda m_{n}(\lambda_{n}^{2}+\Omega_{n}^{2}),
 \label{qos:sc2}
\end{equation}
\begin{equation}
C_{3n}=-m_{n}^{2}D_{qqn}[\lambda_{n}^{2}+\Omega_{n}^{2}
+(\mu_{n}+\lambda_{n})^{2}]-2\,D_{pqn}m_{n}(\mu_{n}+\lambda_{n})-D_{ppn}.,
 \label{qos:sc3}
\end{equation}
\end{subequations}
\end{widetext}
The denominator in the right hand side of Eq.~(\ref{qos:h2}) is
the product of two sums of squares, and hence is guaranteed to be
positive. The numerator is quadratic in $\Delta\omega$ of the form
\begin{eqnarray}
y&=&a \Delta\omega^{2}+b \Delta\omega+c
\nonumber \\
&=&(m_{n}^{2}D_{qqn})\Delta\omega^{2}-m_{n}\lambda_{n}\hbar\Delta\omega
\nonumber \\
&&+m_{n}^{2}(\lambda_{n}+\mu_{n})^{2}D_{qqn}+D_{ppn}+
\nonumber \\
&&2(\lambda_{n}+\mu_{n})m_{n} D_{pqn}. \label{qos:quad}
\end{eqnarray}
If $y$ is positive, then $h_{n,\Delta\omega}$ is positive as well.
The coefficient of the quadratic term is positive, thus $y$ has
positive concavity.  If $y=0$ has no real roots, then $y$ must be
positive, which can be tested by the condition $4ac-b^{2}>0$.
Upon some rearrangement, we can write
\begin{eqnarray}
4ac-b^{2}&=&4m_{n}^{2}\{[m_{n}D_{qqn}(\lambda_{n}+\mu_{n})+D_{pqn}]^{2}
\nonumber\\
&&+D_{qqn}D_{ppn}-D_{pqn}^{2}-(\frac{\lambda_{n}\hbar}{2})^{2}\}.
\label{qos:quad2}
\end{eqnarray}
The first term inside the braces is a square and hence positive,
while the remaining terms satisfy Eq.~(\ref{ep:constraint}) and so
the resulting expression is necessarily positive.  Thus it is
sufficient that our open system model for the oscillators is of
the Lindblad form (as discussed in the previous section) to
guarantee that the rotating wave approximation of the weak
coupling limit generates a Lindblad form for the evolution of the
system.

\section{\label{ex} Example: Oscillator linearly coupled to bath}

In this section we illustrate our results with a test model
consisting of an oscillator linearly coupled to a damped
oscillator bath.  To simplify notation we take $\rho$ to be the
reduced density operator for the system. The system has a nominal
hamiltonian given by
\begin{equation}
H_{S}=\frac{1}{2}m_{S}\omega_{S}^{2}Q^{2}+\frac{P^{2}}{2m_{S}},
 \label{ex:ham}
\end{equation}
and the interaction with the environment is given by
\begin{equation}
U_{I}=\sum_{n}C_{n}Q\otimes q_{n}.
 \label{ex:int}
\end{equation}
With this choice, we can write the operators $V_{n,\Delta\omega}$
in terms of the system creation and annihilation operators
$a^{\dag}$ and $a$
\begin{equation}
V_{n,\Delta\omega}=C_{n}\sqrt{\frac{\hbar}{2m_{S}\omega_{S}}}
\,(a^{\dag}\,\delta_{\Delta\omega,\omega_{S}}
+a\,\delta_{\Delta\omega,-\omega_{S}}).
 \label{ex:V_delta}
\end{equation}
The contribution to the hamiltonian through $\overline{K}$, as it
appears in Eq.~(\ref{qos:avgK6}), becomes
\begin{eqnarray}
\Delta H &=&\frac{1}{\hbar}\sum_{n,\Delta\omega}
(S_{n,\Delta\omega}V_{n,\Delta\omega}
V_{n,\Delta\omega})^{\dag}\nonumber\\
&=&\hbar\delta \omega_{S}(aa^{\dag}-\frac{1}{2})+\Delta E,
\label{ex:dH}
\end{eqnarray}
where
\begin{equation}
\delta \omega_{S}=\sum_{n} \frac{C_{n}^{2}}{2m_{S}\omega_{S}\hbar}
(S_{n,\omega_{S}}+S_{n,-\omega_{S}})
 \label{ex:d_omega}
\end{equation}
and
\begin{equation}
\delta E=\sum_{n} \frac{C_{n}^{2}}{2m_{S}\omega_{S}}
(S_{n,\omega_{S}}-S_{n,-\omega_{S}}).
 \label{ex:d_E}
\end{equation}
Thus $\delta \omega_{S}$ is simply a frequency shift for the
system and $\delta E$ is a C-number shift in the energy. The
remaining contributions to $\overline{K}$ essentially are of the
form of Eq.~(\ref{ep:mesns}) with $\mu = 0$, that is
\begin{eqnarray}
\overline{K}[\rho] &=&-\frac{i}{\hbar}[\hbar\delta
\omega_{S}(aa^{\dag}-\frac{1}{2})+\Delta E,\rho]\nonumber\\
&&-\frac{i}{2\hbar}\lambda([q,\{p,\rho\}]
-[p,\{q,\rho\}])\nonumber\\
&&- \frac{D_{pp}}{\hbar^{2}}[q,[q,\rho]]
-\frac{D_{qq}}{\hbar^{2}}[p,[p,\rho]].
 \label{ex:me1}
\end{eqnarray}
The dissipation and diffusion coefficients are given by
\begin{eqnarray}
\lambda &=&\sum_{n}\frac{C_{n}^{2}}{2\hbar m_{S}\omega_{S}}
(h_{n,\omega_{S}}-h_{n,-\omega_{S}})\nonumber\\
D{pp} &=&(m_{S}\omega_{S})^{2}D{qq} \nonumber\\
&=&\sum_{n}\frac{C_{n}^{2}}{4}
(h_{n,\omega_{S}}+h_{n,-\omega_{S}}) \label{ex:me1coef}
\end{eqnarray}

If the environment degrees of freedom are thermalized (i.e. driven
asymptotically to a Gibbs state) then Eq.~(\ref{ep:gibbs}) holds
and the dissipation and diffusion coefficients are given by
\begin{widetext}
\begin{eqnarray}
\lambda &=&\sum_{n}\frac{C_{n}^{2}\lambda_{n}}{ m_{S}^{2}
(\lambda_{n}^{2}+(\Omega_{n}-\omega_{S})^{2})
(\lambda_{n}^{2}+(\Omega_{n}+\omega_{S})^{2})}
\nonumber\\
D_{pp} &=&\sum_{n}\frac{C_{n}^{2}\hbar \coth{(\frac{\hbar
\omega_n}{2k_{B}T})}
[(\lambda_{n}+\mu_{n})(\lambda_{n}^{2}+\Omega_{n}^{2})
+(\lambda_{n}-\mu_{n})\omega_{S}^{2} ]}{ m_{S}\omega_{n}
(\lambda_{n}^{2}+(\Omega_{n}-\omega_{S})^{2})
(\lambda_{n}^{2}+(\Omega_{n}+\omega_{S})^{2})} \label{ex:megcoef}
\end{eqnarray}
and the frequency shift is given by
\begin{equation}
\delta\omega_{S} = \sum_{n}\frac{C_{n}^{2}
(\lambda_{n}^{2}+\Omega_{n}^{2}-\omega_{S}^{2})}
{2m_{S}^{2}\omega_{S}
(\lambda_{n}^{2}+(\Omega_{n}-\omega_{S})^{2})
(\lambda_{n}^{2}+(\Omega_{n}+\omega_{S})^{2})}
 \label{ex:megd_w}
\end{equation}
\end{widetext}

From Eq.~(\ref{ep:momt1}), the dissipative time scales of the
environment are determined by $\lambda_{n}-\mu_{n}$ and
$\lambda_{n}+\mu_{n}$.  If dissipation is weak, then
\begin{equation}
\lambda_{n}-\mu_{n},\,\lambda_{n}-\mu_{n} \ll \omega_{n}
\end{equation}
and
\begin{equation}
\Omega_{n}\approx\omega_{n}.
\end{equation}
The dissipation and diffusion coefficients each have a factor of
\begin{eqnarray}
&&\frac{1}{(\lambda_{n}^{2}+(\Omega_{n}-\omega_{S})^{2})}
\nonumber\\
&\approx&\frac{1}{(\omega_{n}-\omega_{S})^{2}}
\end{eqnarray}
from which we see that for a weakly damped environment, the
greatest effect is from the oscillators in the environment with
frequencies close to the system's frequency.  With this
approximation, we can further simplify the dissipation and
diffusion coefficients with:
\begin{eqnarray}
\lambda&\approx&\sum_{\omega_{n}\approx\omega_{S}}
\frac{C_{n}^{2}} {4m_{S}^{2}\omega_{S}^{2} \lambda_{n}}
\nonumber\\
D{pp}&\approx&\frac{\hbar m_{S}\omega_{S}
}{2}\coth{\frac{\hbar\omega_{S}}{2k_{B}T}}
\sum_{\omega_{n}\approx\omega_{S}} \frac{C_{n}^{2}}
{4m_{S}^{2}\omega_{S}^{2} \lambda_{n}}\,. \label{ex:megresults}
\end{eqnarray}
Thus the parameters of $\overline{K}$ satisfy Eq.~(\ref{ep:gibbs})
and the system is driven towards thermal equilibrium by an
effective evolution of the Lindblad form.

\section{\label{done} Conclusions}

We have constructed open environment models to account for the
environment's interaction with the ``rest of the universe".  In
Section~\ref{class}, our brief investigation of a classical model
provides some insight and expectations for the development of a
quantum mechanical model.  The review the open system model in
Section~\ref{env_prop} served to establish many properties used in
our derivation of the effective master equation in
Section~\ref{proj}.  Although Section~\ref{proj} largely follows
previous work, the introduction of nonunitary evolution for the
environment provided some novel aspects to the master equation
derivation.  We were able to show that an effective master
equation of the Lindblad form could be obtained for a rich family
of dissipative environment models.  Our illustration of the
resulting master equation with a bilinear environment-system
interaction provided an demonstration in which an environment
dynamically driven toward thermal equilibrium can naturally result
in the dynamical thermalization of the system of interest.

\bibliography{denv}
\end{document}